\documentstyle[12pt]{article}
\begin{document}
\baselineskip=15pt

\newcommand{\be}{\begin{equation}}
\newcommand{\ee}{\end{equation}}
\newcommand{\bq}{\begin{eqnarray}}
\newcommand{\eq}{\end{eqnarray}}
\newcommand{\x}{{\bf x}}
\newcommand{\p}{\varphi}
\newcommand{\Sc}{Schr\"odinger\,}
\newcommand{\del}{\nabla}
\begin{titlepage}
\rightline{DTP 95/39}
\vskip1in
\begin{center}
{\large THE VACUUM FUNCTIONAL AT LARGE DISTANCES}
\end{center}
\vskip1in
\begin{center}
{\large
Paul Mansfield

Department of Mathematical Sciences

University of Durham

South Road

Durham, DH1 3LE, England}

{\it P.R.W.Mansfield@durham.ac.uk}
\end{center}
\vskip1in
\begin{abstract}
\noindent
For fields that vary slowly on the scale of the lightest mass
the logarithm of the vacuum functional can be expanded as a sum of
local functionals, however this does not satisfy the obvious form of the
\Sc equation.
For $\p^4$ theory we
construct the appropriate equation that this expansion does satisfy.
This reduces the eigenvalue problem for the Hamiltonian to a set of
algebraic
equations. We suggest two approaches to their solution. The first
is equivalent
to the usual semi-classical expansion whilst the other is a new
scheme that may also be applied to theories that are classically
massless but in which mass is generated quantum mechanically.
\end{abstract}

\end{titlepage}
%%%%%%%%%%%%%%%%%%%%%%%%%%%%%%%%%%%%%%%%%%%%%%%%%%%%%%%%%%%%%%%%%%%%
%%%%%%%%%%%%%%%%%%%%%%%%%%%%%%%%%%%%%%%%%%%%%%%%%%%%%%%%%%%%%%%%%%%%

For fields that vary slowly on the scale of the lightest mass the
logarithm of  the vacuum functional of a quantum field theory reduces
to a sum of local quantities.
The purpose of this letter is to construct the \Sc equation that acts
directly
on this local expansion.
This
reduces the eigenvalue problem for the Hamiltonian to a set of
algebraic equations for which we suggest two methods of solution.
One of them is the familiar semi-classical expansion, the other
is a new approach that would
apply even when mass is generated quantum mechanically in a theory
that is classically massless.
\bigskip

 To begin with, consider
a free massive scalar field theory in D+1 dimensions. In the \Sc
representation the Hamiltonian
is $H=-{1\over 2}\Delta +\int d^D\x\,{1\over 2}\left( \nabla\p
\cdot\nabla\p +m^2 \p^2\right)$, (see \cite{Jac} for a review of
the \Sc
representation in field theory).
The unregulated Laplacian, $\Delta$, is
$\int d\x\,\delta^2 /\delta\p^2 (\x )$, and the vacuum functional
$\Psi [\p ]\equiv\langle \p |0\rangle=exp\, -\int d^D\x\,
\p\Gamma\p/2\equiv exp\,W_0$ with $\Gamma=+\sqrt{-\nabla^2+m^2}$.
The corresponding energy eigenvalue, E, is proportional to the
functional trace of $\Gamma$ and so diverges. The Laplacian needs
to be regulated so
introduce a momentum cut-off,
$1/s$, by defining
\be
\Delta_s=\int_{p^2<\,1/s} {d^Dp\over (2\pi)^D}{\delta^2\over\delta
\bar\p (p)\delta \bar\p(-p)},
\label{eq:regulo}
\ee
where $\bar\p ({\bf p})=\int d^D\x\,\p (\x)\,exp\,-i{\bf p}\cdot\x$.
The vacuum energy density ${\cal E}=E/V$ is now well-defined and
diverges as the cut-off
is removed

\be
 {\cal E}={1\over 2V}\Delta_s W_0={1\over 2}\int_{p^2<\,1/s}
{d^D p\over
(2\pi)^D}{\sqrt {{\bf p}^2+m^2}}
\sim {k\over (D+1)s^{(D+1)/2}}\quad as\,\,\,s\rightarrow 0
\label{eq:energy}
\ee
Where $k$ is the area of the unit sphere in D dimensions divided by
$2(2\pi)^D$.

\bigskip
If $\p$ varies slowly in space on the scale of $1/m$ then
$W_0[\p]$ simplifies to a local expression obtained by
expanding $\Gamma$ as $m-\nabla^2/(2m)-(\nabla^2)^2/(8m^3)..$ to give
\be
\int d^D\x \left({m\over 2}\p^2+{1\over 4m}(\nabla \p)^2
-{1\over 16 m^3}(\nabla^2\p)^2..\right)\equiv W_0^{loc}
\label{eq:freexp}
\ee
Similar local expansions may be performed for the wave-functionals
given by  \cite{Kief1} \cite{Kief2} describing fields coupled to
gauge-potentials. A local vacuum-functional for Yang-Mills was first
discussed in \cite{Jeff} where it leads heuristically to an area law
for the Wilson loop via a kind of dimensional reduction.   Quark
confinement from dimensional
reduction was discussed in \cite{Olesen}.
Dimensional reduction resulting from the local nature of the \Sc vacuum
functional \cite{Paul} has been generalised by Horiguchi {\it et al}
to the case of
the Wheeler-De Witt equation to look for a new phase in quantum gravity
beyond the Planck scale \cite{Horiguchi}.
We expect that the vacuum functional of the interacting scalar theory
can similarly be expanded as a local quantity for slowly varying $\p$.
The vacuum functional satisfies the \Sc equation, the existence of
which was shown by Symanzik,\cite{Sym}.
In principle this determines the coefficents of the expansion. However
the local expansion does {\it not} satisfy the obvious form of this
equation
because expanding in local quantities does not commute with removing the
cut-off, $s\rightarrow 0$, even for the free theory. To see this
apply the regulated Laplacian $\Delta_s$ directly to the local expansion
(\ref{eq:freexp})
to obtain the energy density as
\be
{1\over 2V}\Delta_s W_0^{loc}=\int_{p^2<\,1/s} {d^Dp\over (2\pi)^D}
\left({m\over 2}-{p^2\over 4m}-{(p^2)^2\over 16m^3}..\right)
=\sum_{n=0}^\infty {\alpha_n \over (m^2s)^{n+D/2}}
\label{eq:enexp}
\ee
(where $\alpha_n=km^{D+1}\Gamma({3\over 2})/ (\Gamma ({3/ 2}-n)
\Gamma (n+1)(D+2n))$).
This expression appears to have divergences of increasing order as
$s\rightarrow 0$ unlike (\ref{eq:energy}) which correctly
gives the behaviour of the vacuum energy as the cut-off is removed.
This is because our local expansion $W_0^{loc}$ is only valid for
slowly varying $\p$, that is for $\bar\p ({\bf p})$ with support
in $p^2<m^2$, hence (\ref{eq:enexp}) is only valid for $sm^2>1$,
i.e. for large $s$.
To obtain information about the vacuum energy for small $s$
from this expansion we must re-sum the series.
First define the continuation of the vacuum energy to complex $s$ by
\be
{\cal E}(s)={1\over s^{D/2}}\int_{p^2<1} {d^Dp\over (2\pi)^4}
\sqrt{m^2+p^2/s}
\label{eq:defin}
\ee
By choosing the square-roots to have cuts on the negative real axis
this is analytic throughout the complex $s$-plane with the
negative real axis removed. For $|s|m^2>1$ it has a
large $s$ expansion which is just (\ref{eq:enexp}).
Let $C$ be a key-hole shaped contour running just under the
negative real axis up to
$s=-s_0$ with $s_0>1/m^2$, around the circle of radius $s_0$ centred
on the origin
and then back to $s=-\infty$ running just above the negative real axis.
The integral
\be
I(\lambda)={1\over 2\pi i}\int_C {ds\over s}{\cal E}(s)\,e^{\lambda s}
\label{eq:I}
\ee
may be evaluated using the large $s$ expansion (\ref{eq:enexp})
as
\be
\sum_{n=0}^\infty {\alpha_n \over \Gamma (n+1+D/2)}\left({\lambda\over
 m^2}\right)^{n+D/2}
\label{eq:resum}
\ee

We can also evaluate the integral by collapsing the contour
$C$ until it just surrounds the negative real axis. By taking $\lambda$
to be real, positive and
very large the integrand will be exponentially suppressed away from
the vicinity of the origin so that the integral is determined by
${\cal E}(s)$ for small s. In fact
$I(\lambda )\sim \lambda^{(D+1)/2}\sum_{n=0}^\infty
\tilde\alpha_n\lambda^{-n}$ as $\lambda\rightarrow\infty$ with constant
$\tilde\alpha_i$ up to exponentially suppressed terms, so that
$I(\lambda)$ provides a regularisation
of the vacuum energy (\ref{eq:energy}). If instead we had been dealing
with a function
${\cal E}(s)$ that was finite in the vicinity of the origin with poles
and
cuts only on the negative real axis  we would
have obtained ${\cal E}(0)$ up to exponentially suppressed terms.
In an interacting theory we would only be able to compute a finite
number of terms of these expansions. Now (\ref{eq:resum}) is an
alternating series so truncating it
at $n=N$ gives an error less than
$|{\alpha_{N+1}\lambda^{N+1+D/2}\over\Gamma(N+2+D/2)m^{2N+2+D}}|$
which for large $N$ behaves as $(e\lambda/
m^2N)^{(N+1+D/2)}/N$. Since we want to take $\lambda$ large set
$\lambda=
N\mu^2$, then the truncation error goes to zero with large $N$
provided the $N$-independent mass-scale $\mu$ is smaller than
the particle mass
$m$. This shows that we can extract information about the
high momentum cut-off theory by working to a finite order, $N$,
with the local expansion of the vacuum functional valid for slowly
varying
fields.

\bigskip
We now turn to the interacting theory. Symanzik \cite{Sym} showed
that for $D=3$ the \Sc functional is finite when the cut-off is
removed provided that in addition to the usual renormalisation
procedure
there is a further field renormalisation to take account of
divergences associated with the boundary. This was taken further in
\cite{Hugh}.
The appropriate renormalisation constants are computable in
perturbation theory, but
since the theory is not asymptotically free these are not reliable
so instead we will work with the 1+1 dimensional theory which is
super-renormalisable.
In the usual treatment without boundaries the only divergent Feynman
diagrams for $\p^4$ theory in 1+1 dimensions are those
in which both ends of a propagator are contracted
at the same point. These may be eliminated by normal ordering the
Hamiltonian with respect to a perturbative vacuum to
obtain a regular operator that does not depend on
a cut-off:
\be
\hat H=\int dx:\left({1\over 2}(\hat\pi(x)^2+\hat\p^\prime(x)^2+
M_r^2\hat\p(x)^2)+{g\over 4!}\hat\p^4\right):
\ee
Here $M_r$ and $g$ are finite, and
the normal-ordering is with respect to oscillators with mass
$M_r$.
There are no further divergences in the \Sc functional
due to the boundaries so in this case there is no further
field renormalisation. (This is thought to apply also in Yang-Mills
theories
\cite{lusch}). Now define an operator
$\hat H_s$ constructed from momentum cut off fields
\bq
\hat H_s=\int dx \left({1\over 2}\left( \hat\pi_s^2+\hat\p_s^{\prime
2}+M^2(s)\hat\p_s^2\right) +{g\over 4!}\hat\p_s^4-{\cal E}_s\right)
\eq
where
\bq
\hat\p_s(x)=\int dy\,{\cal G}_s(x,y)\hat\p (y),\quad
\hat\pi_s(x)=\int dy\,{\cal G}_s(x,y)\hat\pi (y)
\eq
and the kernel
\be
{\cal G}_s(x,x^\prime)=\int_{p^2<\,1/s} {d{ p}\over 2\pi}e^{i{p}
 (x-x^\prime)},
\label{eq:kernel1}
\ee
implements a momentum cut-off.
Writing this in terms of normal-ordered operators enables us to express
$\hat H$ as $\lim_{s\downarrow 0} \hat H_s$ provided that
\be
{\cal E}_s={1\over 2}\int_{p^2<1/s}{dp\over 2\pi}\left(\sqrt{p^2+M_r^2}+
{M^2(s)-M_r^2\over
2\sqrt{p^2+M_r^2}}\right)+{g\over 32}\left(\int{dp\over 2\pi}{1\over
\sqrt{p^2+M^2}}\right)^2
\ee
and
\be
M_r^2=M^2(s)+{g\over 4}\int_{p^2<1/s} {dp\over 2\pi}
{1\over\sqrt{p^2+M_r^2}},
\ee
so that the \Sc equation for the vacuum is just
$\lim_{s\downarrow o}\langle \p|\hat H_s |0\rangle
={ E}\langle\p|0\rangle$ or
\bq
\lim_{s\downarrow 0}&&\left(-{1\over 2}\Delta_s+
\int dx\left({1\over 2}\left(\p_s^\prime(x)^2+ M^2(s)\p_s (x)^2\right)
+{g\over 4!}\p_s(x)^4 -{\cal E}(s)\right)\right)\Psi\nonumber\\
&\equiv& \lim_{s\downarrow 0} \quad (-{1\over 2}\Delta_s+V_s)\Psi
\equiv\,\lim_{s\downarrow 0} \quad H_s\Psi=E\Psi
\label{eq:one}
\eq
If we evaluate this expression for a $\p$ of that has no Fourier modes
with momentum greater than $1/s$ then we can replace $\p_s$
in this expression by $\p$ itself.

\bigskip
  We will now show that the limit of small $s$ may be taken by
re-summing
the large $s$ behaviour, just as for the vacuum energy of the free
theory, enabling us to compute the coefficents
an expansion  of the logarithm of the vacuum functional
in terms of local quantities,
and the eigenvalues of the Hamiltonian.
We begin by showing that $\Delta_s\Psi [\p ]$ extends to an
analytic function in the complex $s$-plane with the negative real
axis removed. Consider $\langle \p|\hat\pi (x,0)
\,\hat\pi(x^\prime,0)|0\rangle$,
written in terms of a Euclidean functional integral. It will be helpful
to make the $\p$ dependence of  $\langle \p|$ explicit.
To do this we can use a bra $\langle D|$ which is annihilated
by $\hat\p$. ($D$ stands for Dirichlet). Then
\be
\langle \p|=\langle D|e^{i\int dx \p(x)\hat\pi (x)},
\ee
so that
\be
{\delta\over\delta\p (x)}\langle \p|=i\langle \p|\hat\pi (x),
\ee
and
using the canonical commutation relations $\langle \p|\hat\p=\p
\langle \p|$. Now in Minkowski space
\be
T\left(\hat\pi (x,t)
\,\hat\pi(x^\prime,t^\prime)\right)=
{\partial^2\over\partial t\,\partial t^\prime}T\left(
\hat\p (x,t)
\,\hat\p(x^\prime,t^\prime)\right)
-i\delta (x-x^\prime)\delta(t-t^\prime)
\label{eq:Tord}
\ee
so that if we use the standard relationship between functional
integrals and time-ordered expectation values we can express
$\langle\p|e^{-it_0\hat H}|D\rangle$ as the functional integral

\be
\int {\cal D}\tilde\p \,e^{iS[\tilde\p]+\int dx (i\p\dot{\tilde\p}+
\Lambda \p^2)}
\ee
where $|D\rangle$ implies Dirichlet boundary conditions $\tilde\p=0$
at times $0$ and $t_0$.
$\Lambda$ is a regularisation of the $\delta (0)$ that arises from
(\ref{eq:Tord}).
Similarly the time ordered Green's function
\be
T\langle \p|e^{-it_0\hat H}\hat\p (x,t)
\,\hat\p(x^\prime,t^\prime)|D\rangle=
\int {\cal D}\tilde\p \,e^{iS[\tilde\p]+\int dx
(i\p\dot{\tilde\p}+\Lambda \p^2)}
{\tilde\p}(x,t){\tilde\p} (x^\prime ,t^\prime),
\ee
Again making use of (\ref{eq:Tord}), rotating to Euclidean space and
taking the limit $t_0\rightarrow \infty$ gives for $0>t\ge t^\prime$

\bq
&&\langle \p|\hat\pi (x,t)
\,\hat\pi(x^\prime,t^\prime)|0\rangle =\nonumber\\
&&-\int
{\cal D}\tilde\p e^{-S_E[\tilde \p ]+\int dx (\p
\dot{\tilde\p}+\Lambda\p^2)}\left(\dot{\tilde\p}
(x,t)\dot{\tilde\p} (x^\prime ,t^\prime)-\delta (x-x^\prime)
\delta(t-t^\prime)
\right)
\label{eq:extdel}
\eq
where the integration variable $\tilde \p$ is now defined on the
Euclidean
half-plane, $t\le 0$, with the boundary condition $\tilde\p (x,0)=0$.
To obtain $\langle \p|\hat\pi (x,0)
\,\hat\pi(x^\prime,0)|0\rangle$ we can either take the limit
as $t,t^\prime \uparrow 0$ keeping $t>t^\prime$, in which case the
$\delta$-
functions will not contribute, or we can take $t=t^\prime=0$
having first cancelled the $\delta$-functions against a divergence coming from
$\dot{\tilde\p} (x,t)\dot{\tilde\p} (x^\prime ,t^\prime)$. A similar
argument applies to the $\Lambda \p^2$ terms. Whichever approach we
take affects only the $\p$-independent and $\int dx\, \p^2$
contributions to the connected
diagrams in $\Delta_s\Psi$.
For example when we integrate against ${\cal G}_s$ to get $\Delta_s\Psi$
the delta-functions give a term proportional to $\Psi/\sqrt s$
which contributes only to the $1/\sqrt s$ part of the vacuum energy in
(\ref{eq:one}).
Apart from this the  $\delta$-functions  have no effect, and henceforth
we
ignore them. Setting $t=t^\prime=0$ in the remaining term, rotate
co-ordinates so that the two points $(x,0)$
and $(x^\prime, 0)$ differ by a Euclidean time $\tau=|x-x^\prime|$
to obtain
\be
\int {\cal D}\tilde\p_r e^{-S_{Er}[\tilde \p ]+\int dt \p
{\tilde\p}_r^\prime}{\tilde\p}^\prime_r
(0,\tau){\tilde\p}^\prime_r (0,0)
\ee
where now the integration variable $\tilde{\p}_r$ is defined on the
half-plane
$x>0$. This can be interpreted as the time-ordered vacuum expectation
value
of fields that evolve in Euclidean time with a Hamiltonian $\tilde H$
defined on the half-line $x>0$
\be
T_E\langle 0_r|e^{\int dt \p(t){\hat\p^\prime (0,t)}}{\hat\p}^\prime
(0,\tau){\hat\p}^\prime (0,0)|0_r\rangle
\label{eq:rotvev}
\ee
The vacuum $|0_r\rangle$ corresponds to the rotated Hamiltonian.
Expanding the exponential and making the Euclidean time dependence
explicit reduces this to a sum of terms of the form
\bq
\int_\tau^\infty dt_n\int_\tau^{t_n}dt_{n-1}..\int^{t_{p+1}}_\tau
dt_p\int^\tau_0dt_{p-1}..\int^{t_{q+1}}_0dt_q\int^0_{-\infty}dt_{q-1}..
\int^{t_2}_{-\infty}dt_1\, \p(t_n)..\p(t_1)&&\nonumber\\
\langle 0_r|\hat\p^\prime e^{-\tilde H(t_n-
t_{n-1})}\hat\p^\prime ..e^{-\tilde H(t_p-
\tau)}\hat\p^\prime e^{-\tilde H(\tau-t_{p-1})}\hat\p^\prime ..
e^{-\tilde Ht_q}\hat\p^\prime e^{\tilde Ht_{q-1}}\hat\p^\prime ..
e^{-\tilde H(t_2-
t_{1})}\hat\p^\prime |0_r\rangle
\eq
The integrals may be computed if we first Fourier decompose $\p(t_i)$ as
\be
\p(t_i)=(2\pi)^{-1}\int dk_i\,
\bar\p (k)exp\,(ik_it_i).
\ee
This gives a delta-function conserving the
total momentum $\sum k_i$ and
for $t_i$ with $i\ge p$ this gives insertions of $(\tilde H
-i\sum k_i)^{-1}$ whereas for $i<q$ it gives insertions of  $(\tilde H
+i\sum k_i)^{-1}$. The remaining $t$-integrals may be done by inserting
a basis of eigenstates of $\tilde H$ between each operator which leads to
a sum of products of energy denominators of the form
$(E_1-E_2-i\sum k_j)^{-1}$
multiplied by exponentials of $\tau$ of the form
$exp\,-(E_1-i\sum k_j)$,
so that the $\tau$ dependence may be made explicit as a sum of
integrals over the spectrum of $\tilde H$
\be
\int dE\,dk_1..dk_n\bar\p(k_1)..\bar\p(k_n)\delta (\sum k_i)
\left(\rho_0+
\rho_1e^{ik\tau}
+\rho_2e^{i(k_1+k_2)\tau}+..\right) e^{-E\tau}
\ee
Integrating this against the kernel ${\cal G}_s$ gives
$\Delta_s\Psi$ as a sum of terms like
\be
\int_{p^2<s^{-1}}{dp\over\pi}\int dE\,dk_1..dk_n\bar\p(k_1)..\bar\p(k_n)
\delta (\sum k_i)
\sum{\rho_j(E,k_1,..,k_n)\over E-i(p+\sum k_i)}
\ee
We will make the dependence on $s$ more explicit by setting
$p=q/\sqrt s$.
Also we will scale the function $\p$ with $s$ by evaluating this
for $\p$ of the form
$\p (x)=f(x/\sqrt s )\equiv\phi_s (x)$ so that $\bar\p (k)=
\bar f(k\sqrt s )\sqrt s$, and we take $\bar f(k)$ to vanish outside
$|k|<\kappa<<1 $.
This enables us to scale the $k_i$ integrals
to obtain
\be
\int_{q^2<1}{dq\over\pi}\int dE\,dk_1..dk_n\bar f(k_1)..\bar f(k_n)
\sqrt s\delta (\sum k_i)
\sum{\rho_j(E,k_1,..,k_n)\over \sqrt s E-i(q+\sum k_i)}
\label{eq:rhoexpan}
\ee
The functions $\rho_j$ acquire a dependence on $s$ via the energy
denominators that can be written after scaleing of $k_i$ as
$1/((E_1-E_2)-i(\sum k)/\sqrt s)$. Up to this point $s$
has been real and positive but the expressions we obtain for
$(\Delta_s\Psi)[\phi_s]$
can now be continued as analytic functions to the whole complex
$s$-plane
excluding the negative real axis, (at least order by order in the
expansion
in powers of $\phi_s$ that we used to treat the exponential in
(\ref{eq:rotvev})). Similarly $((H_s-E)\Psi)[\phi_s]$  extends to
an analytic function of $s$ with singularities only on the negative real
axis, so its finite value at $s=0$ can be obtained from an integral over
a contour on which $|s|$ is arbitrarily large. Thus the \Sc
equation becomes
\be
\lim_{\lambda\rightarrow\infty}{1\over 2\pi i}
\int_C {ds\over s}\,e^{\lambda s}((H_s-E)\Psi)[\phi_s]=0
\label{eq:IJ}
\ee
We take $C$ as before.
Setting $\Psi=e^W$ gives
\be
\lim_{\lambda\rightarrow\infty}{1\over 2\pi i}
\int_C {ds\over s}\,e^{\lambda s}\left(-{1\over 2}\left(\Delta_sW
+\int dxdy{\cal G}_s(x,y){\delta W\over\delta\p(x)}
{\delta W\over\delta\p(y)}\right)-V_s-E\right)[\phi_s]=0
\label{eq:II}
\ee
where we have divided through by $\Psi [\p_s]$ which has a finite
limit as $s\downarrow 0$ as can be checked in perturbation theory.
Since $\kappa<<1$ we can replace ${\cal G}_s(x,y)$ in (\ref{eq:II})
by a delta-function. For a general $W[\p]$ we would expect to be able
to expand in powers of $\p$ to give
\be
W[\p]=\sum_{n=1}^\infty\int {dk_1\over 2\pi}..
{dk_{2n}\over 2\pi}\bar\p(k_1)..
\bar\p(k_{2n})\delta(\sum k_i)\Gamma_n (k_1,..,k_{2n}).
\ee
An expansion in terms of local quantities
would result from expanding $\Gamma_n$ in positive powers
of momenta for small momenta. Applying $\Delta_s$ to $W[\p]$
gives
\bq
&&\Delta_s W[\p]=\nonumber\\
&&\int_{p^2<{1\over s}}{dp\over 2\pi}
\sum_{n=1}^\infty n(n-1)\int {dk_3\over 2\pi}..
{dk_{2n}\over 2\pi}\bar\p(k_3)..
\bar\p(k_{2n})\delta(\sum k_i)\Gamma_n (p,-p, k_3,..,k_{2n}).
\eq
For large $s$ the momentum $p$ is small so we can expand $\Gamma_n
(p,-p, k_3,..,k_{2n})$ in positive powers of $p$. If we evaluate
$\Delta_s W$
for $\p=\phi_s$ then we can also expand in positive
powers of the other momenta $k_i$, again obtaining a local expression,
since the Fourier transform of $\phi_s$ has
only low momentum modes for large enough $s$.
This
coincides with what we would have obtained by applying $\Delta_s$
directly to the local expansion of $W$ followed by substituting
$\phi_s$ for $\p$. This leads to
a set of algebraic equations which re-formulate the eigenvalue problem
for the Hamiltonian.
Concretely, we study
\be
{1\over 2\pi i}\int_C {ds}\,s^{-1}e^{\lambda s}\left(-{1\over 2}
\left(\Delta_sW_\lambda
+\int dx\left({\delta W_\lambda\over\delta\p}\right)^2\right)-
V_s-E_\lambda\right)[\phi_s]=0
\label{eq:our}
\ee
for large $\lambda$
taking $W_\lambda$ to be local.

\bigskip
        Integration by parts gives linear relations between local
expressions so we pick as a  linearly independent basis
expressions of the form
$\int dx \p^{v_0}(\p^\prime)^{v_1}..(\p^{(n)})^{v_n}$ where $v_n$,
the power of the highest derivative in the expression, must be at
least two. So we take as our local expansion of $W_\lambda$
\bq
W_\lambda=\int dx\Biggl(\p^2(b_0+c_0\p^2)+\p^{\prime 2}(b_1+c_1\p^2+
d_1\p^{\prime 2})&&\nonumber\\
+\sum_2^\infty \p^{(n)2}(b_n+c_n\p^2+d_n\p^{\prime 2}+e_n\p\p^
{\prime\prime}+..)&&\nonumber\\
+\p^4(f_0\p^2+f_1\p^{\prime 2}+f_2\p^{\prime\prime 2}+..)+..&&\Biggr)
\eq
The coefficents $b_i,c_i,..$ depend on $\lambda$ through
(\ref{eq:our}) but since $\Psi$ is finite they are themselves
finite as $\lambda\rightarrow\infty$.
Substituting this into our \Sc equation (\ref{eq:our}) yields
\bq
\int dx\Biggl(&&2\bar{\cal E}(\lambda)+2{\cal E}+
{\sqrt\lambda\over\sqrt\pi^3}\left(4b_0+{8b_1\lambda\over 9}
+{16b_2\lambda^2\over 75}+..+{2b_n\sqrt\pi\lambda^n
\over \Gamma (n+3/2)(2n+1)}+..\right)\nonumber\\
&&+\p^2\left (-\bar M^2(\lambda)+{4b_0^2\over\sqrt\pi}
+{\sqrt\lambda\over\pi}\left(12 c_0+{2c_1\lambda\over 3}
+{c_2\lambda^2\over 5}+..\right)\right)\nonumber\\
&&\p^4\left(-{2g\over 4!\sqrt\pi}+{16b_0c_0\over\sqrt\pi}
+{\sqrt\lambda\over\pi}\left(30 f_0+{2f_1\lambda\over 3}
+{f_2\lambda^2\over 2}+..\right)\right)\nonumber\\
&&+\p^{\prime 2}\left(-{2\over\sqrt\pi}+{16b_0b_1\over\sqrt\pi}
+{\sqrt\lambda\over\pi}\left(2 c_1+2d_1\lambda+
{4c_2\lambda\over 3}
+{d_2\lambda^2\over 3}-{e_2\lambda^2\over 5}+..\right)\right)
\nonumber\\
&&+\p^2\p^{\prime 2}\left({32b_0c_1+96b_1c_0\over\sqrt\pi}
+{12f_1\sqrt\lambda\over\pi}+..\right)\nonumber\\
&&+\p^{\prime \prime 2}\left({32b_0b_2+16b_1^2\over3\sqrt\pi}
+{\sqrt\lambda\over\pi}\left( c_2+{d_1\lambda\over 9}
-{e_2\lambda\over 3}+..\right)\right)\nonumber\\
&&+..\nonumber\\
&&+(\p^{(n)})^2\left({\sum 4b_mb_{n-m}\over \Gamma (n+1/2)}
+{\sqrt\lambda\over\pi}\left({2 c_n\over n!}+{2(d_n-e_n)\lambda\over
3(n+1)!}+..\right)\right)\nonumber\\
&&\quad\quad\quad\quad\quad\quad\quad\quad\quad\quad
\quad\quad +...\quad\quad\quad\quad\quad\quad\quad\quad
\quad\quad\quad\Biggr)=0
\label{eq:ourstoo}
\eq
where $\p=\phi_s$ for $s=1/\lambda$
Requiring the coefficent of each linearly independent function of
$\p$ to vanish gives an infinite set of algebraic
equations. These may be solved in the usual semi-classical
approach by first ignoring the
power series in $\lambda$ so that to leading order
$b_0=\pi^{1/4}\bar M/2,\, c_0=g/(192b_0),\,b_1=1/(8b_0),..$ etc.
This is the same as ignoring $\Delta_sW$ in (\ref{eq:our})
and solving the resulting Hamilton-Jacobi equation as a local
expansion. This local expansion is possible because the full solution
to the Hamilton-Jacobi equation is just the Euclidean action
evaluated on-shell and the classical theory is massive.
Corrections to the coefficents can now be obtained iteratively
by substituting the leading order solutions into $\Delta_sW$.

\bigskip
        For theories
such as Yang-Mills which are classically massless
the classical action does not have a local expansion, so that
the semi-classical approach to solving (\ref{eq:our})
is not viable. However the full theory is believed to be massive,
\cite{Feyn}, so that our expansions
make sense, but we need a different approach to solving the resulting
algebraic equations. We now propose a method in which the expansions
in $\lambda$ are truncated at a certain order and the coefficents
associated with the higher orders estimated using (\ref{eq:rhoexpan}).
Consider, for example,
the $\p$-independent contribution to $\Delta_s W$.
Using (\ref{eq:rhoexpan})
\bq
&&{\sqrt\lambda\over\sqrt\pi^3}\left(4b_0+{8b_1\lambda\over 9}
+{16b_2\lambda^2\over 75}+..+{2b_n\sqrt\pi\lambda^n
\over \Gamma (n+3/2)(2n+1)}+..\right)\equiv \sum \lambda^{n+1/2}\xi_n
\label{eq:alt}\nonumber\\
&&={1\over 2\pi i}
\int_C {ds\over s}\,e^{\lambda s}
\int d\mu(E)\int_{q^2<1} dq {1\over \sqrt s E-iq}
|\langle 0_r|{\hat\p}^\prime
(0)|E\rangle|^2\label{eq:energyint}
\eq
(In the derivation of this formula we dropped the terms due to the
delta-functions in (\ref{eq:extdel}) which only contribute a term
proportional to $1/\sqrt s$ to this expression. Since this has no
effect on the
terms of higher order in $s$ we ignore it here too).
If we expand $(\sqrt s E-iq)^{-1}$ for large $s$ in powers of
$1/\sqrt s$
then for the high order terms the integral over the spectrum is
suppressed by
a large power of $1/E$ so that the dominant contribution will
come from the lowest values of E. These correspond to the lightest
particle
one-particle states for which
\be
d\mu (E)|\langle 0_r|{\hat\p}^\prime
(0)|E\rangle|^2={dp\over E}p^2|\langle 0_r|{\hat\p}
(0)|E\rangle|^2,\quad p=\sqrt{E^2-m^2}.
\ee
For large $n$
\be
\int{dp\over E^{2n+2}}p^2 f(E)=\int dp\,p^2e^{-(n+1)ln(p^2+m^2)}f(E)
\approx \int dp\,p^2e^{-(n+1)p^2/m^2}m^{-2(n+1)}f(m)
\ee
This, together with Stirling's formula gives for large $n$
\be
\xi_n\sim (-)^n{\xi\over n^{3}}\left({e\lambda\over n m^2}\right)^
{n+1/2}\left( 1+O\left({1\over n}\right)\right),\ee
with $\xi$ independent of $\lambda$ and $n$ and proportional to
$m|\langle 0_r|{\hat\p}
(0)|E=m\rangle|^2$.
This shows that at high orders (\ref{eq:alt}) is an alternating series.
The error in truncating
it at order
$\lambda^{N+1/2}$ is proportional to $({e\lambda /
N m^2})^{N+3/2}/N^{3}$.
If we set $\lambda=N\mu^2/e$ then this error goes to zero with
increasing $N$ provided the $N$-independent mass-scale $\mu$
satisfies $\mu <m$, and at the same time $\lambda$ increases as we
increase $N$. Similar, but more complicated calculations show that
the coefficents of $\p^2,\,\p^4,\p^{\prime 2}$ behave like
\be
\sum{ (-1)^n\over {\sqrt n}^{3}}\left(
{e\lambda\over n m^2 }\right)^{n+1/2},\quad
\sum{ (-1)^n\over  n}\left({e\lambda\over n m^2 }\right)^{n+1/2},\quad
\sum{ (-1)^n\over\sqrt n}\left({e\lambda\over n m^2 }\right)^{n+1/2},
\label{eq:asymcoeff}
\ee
so that again if we truncate these series at order $\lambda^{N+1/2}$
the error will go to zero as $N$ is increased. Furthermore, we obtain
asymptotic values of the coefficents for large $n$, e.g.
\be
b_n\sim \left(-{1\over m^2}\right)^n{\bar b\over n^{3/2}},\quad
c_n\sim \left(-{1\over m^2}\right)^n{\bar c\over n^{1/2}},\quad
d_n-e_n\sim \left(-{1\over m^2}\right)^n\bar d n^{3/2}
\ee
with constant $\bar b,\bar c,\bar d$. This suggests the following scheme
for solving our local expansion of the \Sc equation. First truncate
the coefficents of each linearly independent local functional of $\p$
at, say, order $\lambda^{N+1/2}$ taking $\lambda=N\mu^2/e$. This gives
an infinite set of
equations each involving a finite number of unknowns, but with
an infinite number of unknowns overall. These can be solved with
judicious
use of (\ref{eq:asymcoeff}) to estimate the discarded coefficents
taking care to solve those equations for which the truncation error is
smallest.
In this approach there is no expansion parameter, but the approximation
consists of working with large, but finite, values of $N$ and $\lambda$
rather
than the infinite values that occur in (\ref{eq:our}). For the purpose
of illustration only we take the unrealistically small
value $N=0$. Equating to zero the coefficents of $1,\,\p^2,\,
\p^{\prime 2},
\,\p^4,\,\p^2\p^{\prime 2},\p^{\prime\prime 2}$ keeping only the terms
up
to order $\sqrt \lambda$ gives six equations in the unknowns ${\cal E},\,
b_0,\,b_1,\,b_2,\,c_0,\,c_1,\,f_0,\, f_1$. Our asymptotic formulae
for the coefficents associated with higher orders in $\lambda$
are only valid for large $n$ but using them here for
small  $n$ (after reinstating the factors of $(n+1)/n$ that for
large $n$ were
replaced by $1$) we obtain
$c_1=-3\sqrt 2 c_0/m^2,\,f_1=-15f_0/m^2, b_2=-2\sqrt 2 b_1/
(3\sqrt 3 m^2)$.
This gives us six equations in the six unknowns  ${\cal E},\,
b_0,\,b_1,\,c_0,\,f_0,\, m$. Thus we obtain
\be
c_0={\pi\over 12\sqrt\lambda}\left(\bar M^2-{4b_0^2\over\sqrt\pi}
\right),
\quad b_1={1\over 8b_0}\left(1+{ \sqrt\pi\bar M^2-4b_0^2\over 4\sqrt 2
m^2}\right)
\ee
where $b_0$ satisfies
\be
gb_0\sqrt\lambda-\pi\left(\bar M^2-{4 b_0^2\over\sqrt\pi}\right)
\left(b_0^2(16-17\sqrt 2)
+2m^2+{\sqrt{2\pi}{\bar M}^2\over 4}\right)=0
\ee
Note that when the interaction is switched off, i.e. when
$g=0$, we obtain the free solution from $\bar M^2-4b_0^2/\sqrt\pi=0$,
as we should. Clearly it  would be necessary to resort to numerical
techniques to make progress with these algebraic equations.

\bigskip

      We can extend this approach to the calculation of the particle
spectrum. Take the one-particle state corresponding to the lightest
particle at rest to have the form of a pre-factor $P$ mutliplied by
$\Psi$. For
slowly varying fields this pre-factor will be the integral of a local
function
\be
P=\int dx\left(\p+\bar a_1\p^3+\bar a_2\p^5+..+\bar b_1\p\p^{\prime
2}+\bar b_2\p\p^{\prime 4}+..+\bar c_1 \p^3\p^{\prime 2}+..\right)
\ee
The \Sc equation equation for this state reduces to an equation
linear in
$P$
\be
{1\over 2\pi i}\int_C {ds}\,s^{-1}e^{\lambda s}\left( {1\over 2}
(\Delta_s) P+\int dx{\delta P\over\delta\p (x)}
{\delta W\over\delta\p (x)}+mP\right)[\phi_s]=0
\ee
Again this may be reduced to a finite number of equations for a
finite
number of unknown coefficents by truncation and using
formulae for the asymptotic behaviour of those coefficents
derived from the study of $\Delta_sP$ for large $s$.
\bigskip

To summarise:
the logarithm of the vacuum functional reduces to a local
functional for fields that vary slowly on the scale of the
lightest mass
in the theory, but this does not satisfy the obvious form
of the \Sc equation
because expanding in slowly varying fields does not commute
with removing
a short distance cut-off. Instead we  constructed the
appropriate form of the equation for the case of $\p^4$ theory
by re-summing the cut-off dependence
using the analyticity properties of the Laplacian applied to the vacuum.
This \Sc equation reduces the eigenvalue problem for the
Hamiltonian to an infinite set
of algebraic equations. We indicated how these may be solved in
the usual
semi-classical approximation and  suggested a new approximation
scheme
that would also be viable for theories that are classically
massless
but acquire mass quantum mechanically such as the O(N) sigma
model and Yang-Mills theories.

\bigskip
\noindent
Finally, I would like to acknowledge useful conversations
with Hugh Osborn
concerning the \Sc representation in field theory, and a grant
from the
Nuffield Foundation.

%%%%%%%%%%%%%%%%%%%%%%%%%%%%%%%%%%%%%%%%%%%%%%%%%%%%%%%%%%%%%%%


\begin{thebibliography}{88}

\bibitem{Jac} R. Jackiw, Analysis on infinite dimensional
manifolds:
\Sc representation for quantized fields, Brazil Summer
School 1989

\bibitem{Kief1} C. Kiefer, Phys. Rev. D 45 (1992) 2044

\bibitem{Kief2} C. Kiefer and Andreas Wipf,
Ann. Phys. 236 (1994)241

\bibitem{Jeff} J. Greensite,  Nucl.Phys. B158 (1979) 469,
Nucl.Phys. B166 (1980) 113


\bibitem{Olesen} J. Ambjorn, P. Olesen, C. Petersen,
Nucl. Phys. B240 (1984) 189

\bibitem{Paul} P. Mansfield, Nucl. Phys. B418 (1994) 113






\bibitem{Horiguchi} T. Horiguchi, KIFR-94-01, KIFR-94-02,
KIFR-94-03, KIFR-95-01, KIFR-95-02
T.Horiguchi, K. Maeda, M. Sakamoto, Phys.Lett. B344 (1994) 105




\bibitem{Sym} K. Symanzik, Nucl. Phys. B190[FS3] (1983) 1


\bibitem{Hugh} D.M. McAvity and H. Osborn,
Nucl. Phys. B394 (1993) 728

\bibitem{lusch} M. L\"uscher, R. Narayanan, P. Weisz and U. Wolff,
Nucl. Phys. B384 (1992) 168

\bibitem{Feyn} R.P. Feynman Nucl. Phys. B188(1981) 479



\end{thebibliography}
\end{document}